\documentclass[preprint,showpacs,amsmath,amssymb]{revtex4}
\usepackage{amsmath}
\usepackage{graphicx}
\usepackage{amsfonts}
\usepackage{dcolumn}
\usepackage{bm}
\usepackage{ulem}

\usepackage{CJK}

\begin{document}
\title{Complex activated transition in a system of two coupled bistable oscillators}

\author{Hanshuang Chen$^1$} \email{chenhshf@mail.ustc.edu.cn}

\author{Feng Huang$^{2}$}

\author{Chuansheng Shen$^{3}$}

\author{Gang He$^{1}$}

\author{Zhonghuai Hou$^4$} 

\affiliation{ $^{1}$School of Physics and Materials Science, Anhui
University, Hefei, 230039, People's Republic of China
\\$^2$Department of Mathematics and Physics, Anhui
Jianzhu University, Hefei 230601, People's Republic of China
\\$^3$Department of Physics, Anqing Normal University, Anqing, 246011,
China \\$^4$Hefei National Laboratory for Physical Sciences at
Microscales \& Department of Chemical Physics, University of Science
and Technology of China, Hefei, 230026, China }

\date{\today}

\begin{abstract}
We study the fluctuation-activated transition process in a system of
two coupled bistable oscillators, in which each oscillator is driven
by one constant force and an independent Gaussian white noise. The
transition pathway has been identified and the transition rate has
been computed as the coupling strength $\mu$ and the mismatch
$\sigma$ in the force constants are varied. For identical
oscillators ($\sigma=0$), the transition undergoes a change from a
two-step process with two candidate pathways to a one-step process
with also two candidate pathways to a one-step process with a single
pathway as $\mu$ is increased. For nonidentical oscillators
($\sigma\neq0$), a novel transition emerges that is a mixture of a
two-step pathway and a one-step pathway. Interestingly, we find that
the total transition rate depends nonmonotonically on $\mu$: a
maximal rate appears in an intermediate magnitude of $\mu$.
Moreover, in the presence of weak coupling the rate also exhibits an
unexpected maximum as a function of $\sigma$. The results are in an
excellent agreement with our numerical simulations by forward flux
sampling.
\end{abstract}
 \maketitle

\section{Introduction} Fluctuation-activated transition between coexisting stable
states underlies many important physical, chemical, biological and
social phenomena. Examples include diffusion in solids, switching in
nanomagnets \cite{PhysRevB.55.11552} and Josephson junctions
\cite{PhysRevB.9.4760}, nucleation \cite{Kashchiev2000,
Frenkel2004}, chemical reactions \cite{Gillespie1977, Kampen1992},
protein folding \cite{Karplus1994, White1999, Wales2003} and
epidemics \cite{PhysRevLett.97.200602, PhysRevLett.101.078101}. A
detailed theory of transition rates was first developed by Kramers
in 1940 for systems close to thermal equilibrium \cite{Kramers1940}.
wherein the transition rate is determined by the free energy barrier
between the states. Consequently, many generalizations of Kramers'
theory have been widely exploited. For a comprehensive review see
Ref.\cite{RevModPhys.62.251}. Nowadays, these theories have been
commonly utilized for a great many applications in diverse fields
\cite{PhysRevLett.77.783,PhysRevE.62.927,EPL2004,NatureDykman}.

In recent years, there is growing interest in the study of the
activated transition in spatially extended systems with two or more
coupled subsystems. Each subsystem has more than one stable state or
conformation. This is because that many natural and artificial
systems can be viewed as a coarse representation of coupled
subsystems, like arrays of Josephson junctions
\cite{PhysRevLett.76.404}, the power grid \cite{Motter2013}, neural
and gene regulatory networks \cite{PNAS04004341, PNAS06008372,
Plos09004872}, and metapopulations \cite{PhysRevLett.109.138104,
PhysRevLett.109.248102, PhysRevLett.112.148101}. In this context,
two important questions should be answered: what is the rate of the
transition, and how does the transition happen? Generally speaking,
the transition typically follows one of two ways. In the weak
coupling, the transitions in the subsystems happen in a serial way.
While for strong coupling, the transitions in the subsystems become
synchronized and the whole system behaves in a coherent way.
Interestingly, a nontrivial phenomenon, i.e., the transition rate
depends nonmonotonically on coupling between subsystems, was found
in several different scenarios, such as extinction risk
\cite{PhysRevLett.109.138104} and mean fixation time
\cite{PhysRevLett.112.148101} of migrated metapopulations,
nucleation of Ising model \cite{PhysRevE.83.046124} and information
diffusion \cite{arXiv2014} on modular networks.

Inspired by the findings, in this paper we want to use a simple
model to make a systematic investigation for the fluctuation-driven
transition in coupled systems. To the end, we employ a system of two
coupled bistable oscillators where each oscillator is injected to
one constant force and an independent Gaussian white noise. By
varying the coupling strength and the mismatch in the force
constants, we consider the pathway and the rate of the transition,
both by theory and by a rare-event simulation. We find that the
transition process exhibits diverse pathways in different parametric
regions that can include multiple transition pathways and multi-step
transition processes. The transition rate also behave a nontrivial
dependence on the coupling and the mismatch. In particular, there
exists a maximal rate at an intermediate magnitude of coupling
strength. Also, for a weak coupling the rate peaks at a proper force
mismatch.

\section{Model}
We consider a system of two mutually coupled bistable overdamped
oscillators which are forced by statistically independent noises and
constant forces. The system under consideration is governed by the
following stochastic differential equations,
\begin{eqnarray}
\left\{ \begin{array}{l}
{{\dot x}_1} = {x_1} - x_1^3 + {\epsilon _1} + \mu ({x_2} - {x_1})+ \sqrt{2D} {\xi _1}(t) \\
{{\dot x}_2} = {x_2} - x_2^3 + {\epsilon _2} +  \mu ({x_1} - {x_2})+
\sqrt{2D}  {\xi _2}(t)
\end{array} \right.\label{eq1}
\end{eqnarray}
where $\epsilon_{1(2)}$ is the external force constant in the
subsystem 1(2), $\mu$ is the coupling strength, and $D$ is the
intensity of the Gaussian white noises with $\left\langle {{\xi
_i}(t)} \right\rangle  = 0$ and $\left\langle {{\xi _i}(t){\xi
_j}(t')} \right\rangle  = {\delta _{ij}}\delta (t - t')$
($i,j=1,2$). In this paper, we set $\epsilon_1=0.1-\sigma$ and
$\epsilon_2=0.1+\sigma$, where ${\sigma} \in [0,0.1]$ measures the
difference between the external forces acted on the two oscillators.
For $\sigma=0$, the two oscillators are identical; otherwise they
are nonidentical. Initially, we place both the two oscillators on
the left potential wells located on $x_{1(2)}\simeq -1$, and study
the transition process from this metastable state to the most stable
state in which the two oscillators are both right potential wells
near $x_{1(2)}\simeq 1$ in the presence of weak noises. Here, we are
interested in how the coupling strength $\mu$ and the mismatch
$\sigma$ in external forces affect the pathway and rate of the
transition.

\section{Results} To proceed our theoretical analysis, we rewritten Eq.(1) as
\begin{eqnarray}
\dot {\vec x }=  - \nabla V(\vec x) + \sqrt {2D} \vec \xi (t)
\label{eq2}
\end{eqnarray}
where $\vec x=(x_1, x_2)$ and $\vec \xi (t) = ({\xi _1}(t),{\xi
_2}(t))$ are the two-dimensional state variable and noise,
respectively. $V(\vec x)$ is the effective potential that can be
expressed as
\begin{eqnarray}
V({x_1},{x_2}) = \sum\limits_{i = 1}^2 {\left( {-\frac{{1}}{2}x_i^2
+ \frac{1}{4}x_i^4 - {\epsilon _i}{x_i}} \right)}  + \frac{\mu}{2}
{(x_1-x_2)^2}. \label{eq3}
\end{eqnarray}

To give the stationary solutions of the system in the absence of
noises, we numerically solve the equation $\nabla V(\vec x)=0$. The
solutions are classified into three types according to the
stabilities of the solutions: stable node points (potential minima),
saddle points (transition states), and unstable node points
(potential maxima). To distinguish among them, one need to calculate
the eigenvalues of the so-called Hessian matrix $J_{ij}=\partial^2
V/(\partial x_i \partial x_j)$ ($i,j=1,2$). If the two eigenvalues
are both positive (negative), the solutions are (un)stable node
points. If the signs of the two eigenvalues are opposite, the
solutions are saddle points.

\begin{figure}
\begin{center}
\includegraphics [width=10cm]{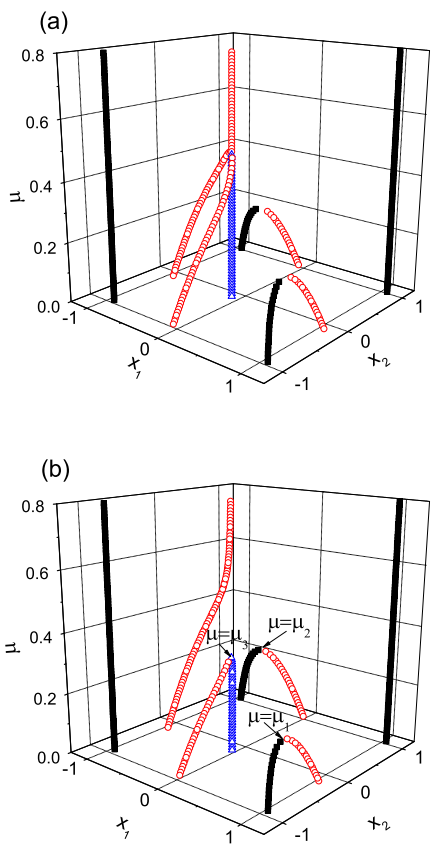}
\caption{(color online) The stationary solutions of Eq.(1) in the
absence of noises for $\sigma=0.0$ (a) and $\sigma=0.05$ (b). Stale
node points, saddle points, and unstable node points are marked by
solid squares, empty circles, and empty triangles, respectively.
\label{fig1}}
\end{center}
\end{figure}

The results for two different typical cases: $\sigma=0$ and
$\sigma=0.05 \neq 0$ are shown in Fig.\ref{fig1}. For $\sigma=0$,
there are nine solutions when the coupling strength $\mu$ is
sufficiently small, denoted by $\vec x_*^{(i)}$ with $i = 1, \ldots
,9$. Four of them, $\vec x_*^{(i)}(i = 1,2,3,4)$, are stable node
points, corresponding to four stable states where both the two
oscillators locate at one of two potential wells. Specifically,
$\vec x_*^{(1)}$ and $\vec x_*^{(2)}$ represent that both the two
oscillators locate at left potential wells and right potential
wells, respectively. $\vec x_*^{(3)}$ represents that the first
oscillator locate at right potential well and the second one locate
at left potential well. And $\vec x_*^{(4)}$ represents that the
first oscillator locate at left potential well and the second one
locate at right potential well. There are four saddle points, $\vec
x_*^{(i)}(i = 5,6,7,8)$ that are transition states for connecting
neighboring stable states. The remaining one point $\vec x_*^{(9)}$
is unstable node whose location lies in two potential barrier of the
two oscillators. With the increment of $\mu$, the stable node point
$\vec x_*^{(3)}$ and the saddle point that connects $\vec x_*^{(3)}$
and $\vec x_*^{(2)}$ approach each other, and collide and annihilate
at $\mu=\mu_1$. Simultaneously, $\vec x_*^{(4)}$ and the saddle
point that connects $\vec x_*^{(4)}$ and $\vec x_*^{(2)}$ approach
each other till they collide and annihilate at $\mu=\mu_2$.
Interestingly, we find that $\mu_1=\mu_2$ if $\sigma=0$ and
$\mu_1<\mu_2$ if $\sigma\neq0$. Thus, for $\sigma=0$ the number
$n_s$ of solutions decreases to $n_s=5$ from $n_s=9$ when $\mu$
passes $\mu_1(=\mu_2)$. While for $\sigma\neq0$, $n_s$ changes from
9 to 7 at $\mu=\mu_1$ and then to 5 at $\mu=\mu_2$. Meanwhile, as
$\mu$ is further increased the saddle point connecting $\vec
x_*^{(1)}$ and $\vec x_*^{(3)}$ and the unstable node point $\vec
x_*^{(9)}$ approach each other, and cease to exist at
$\mu=\mu_3>\mu_2$, such that $n_s$ changes from 5 to 3 at
$\mu=\mu_3$.

\begin{figure}
\begin{center}
\includegraphics [width=10cm]{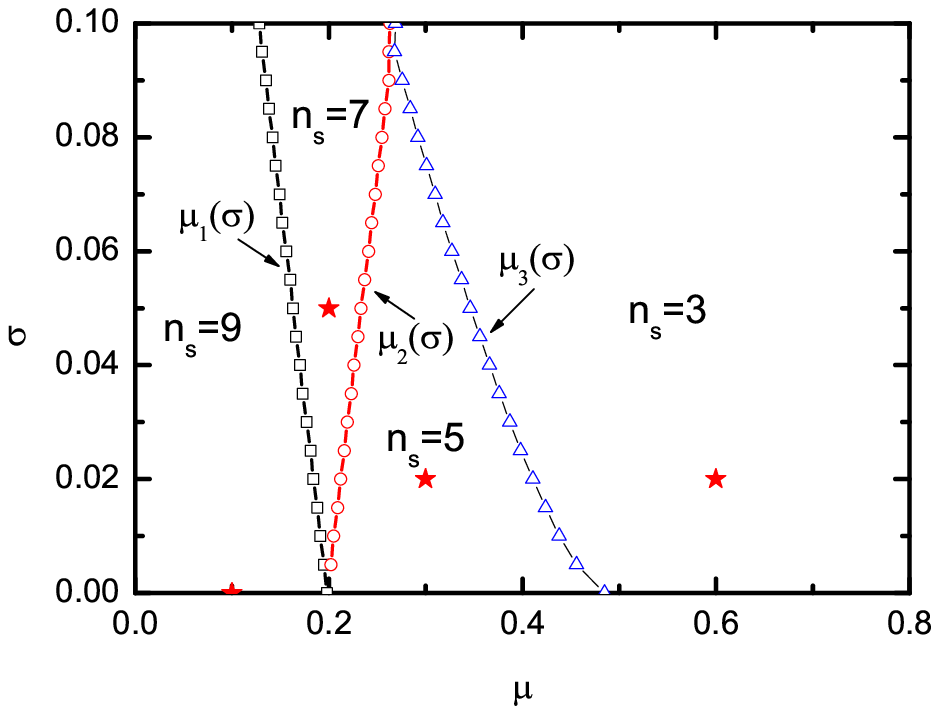}
\caption{(color online) The property of solutions in the
$\mu-\sigma$ plane. The plane are divided into four different
regions according to the number of solutions, and boundary lines
among neighboring regions are indicated by $\mu_1(\sigma)$,
$\mu_2(\sigma)$, $\mu_3(\sigma)$, respectively. \label{fig2}}
\end{center}
\end{figure}

To get a global view, in Fig.\ref{fig2} we plot the phase diagram in
the $\mu-\sigma$ plane. The plane are divided into four different
regions according to the number of solutions. As mentioned above,
$\mu_1(\sigma)$, $\mu_2(\sigma)$, and $\mu_3(\sigma)$ are the
separatrix between the region $n_s=9$ and the region $n_s=7$,
between the region $n_s=7$ and the region $n_s=5$, and between the
region $n_s=7$ and the region $n_s=3$, respectively. Both the region
$n_s=7$ and the region $n_s=5$ have the shape of tongue. With
decreasing $\sigma$ the region $n_s=7$ shrinks until it vanishes
when the lines $\mu_1(\sigma)$ and $\mu_2(\sigma)$ get across at
$\mu=0.195$ and $\sigma=0$. As $\sigma$ is increased, the region
$n_s=5$ is reduced. For $\sigma=0.1$ the lines $\mu_2(\sigma)$ and
$\mu_3(\sigma)$ are very close each other.

\begin{figure*}
\begin{center}
\includegraphics [width=15cm]{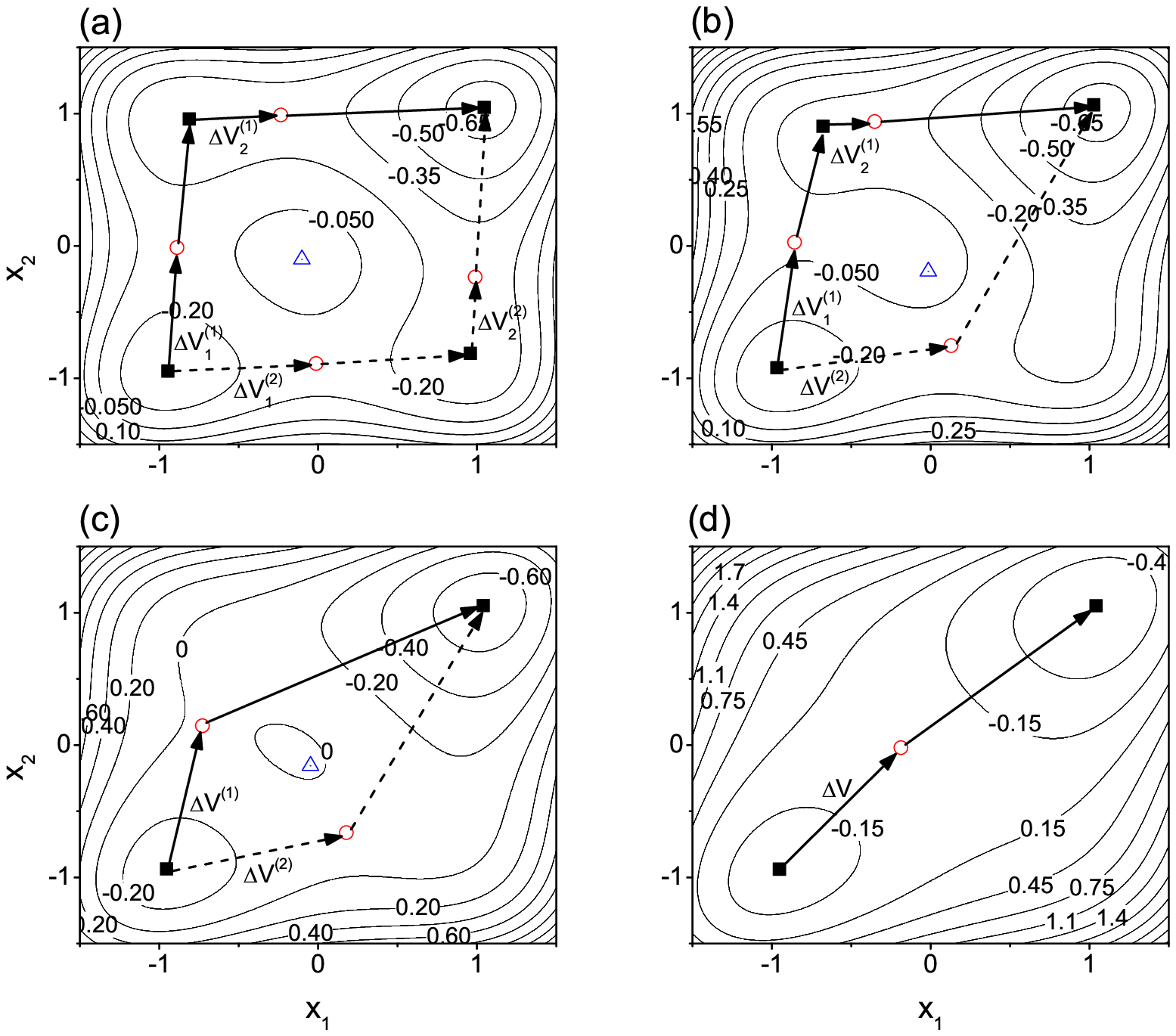}
\caption{(color online) Contour plots of the effective potential $V$
for four representative points marked by stars in Fig.2: $(\mu,
\sigma)=(0.1, 0)$ (a), $(0.2, 0.05)$ (b), $(0.3, 0.02)$ (c), and
$(0.6, 0.02)$ (d). Stale node points, saddle points, and unstable
node points are marked by solid squares, empty circles, and empty
triangles, respectively. The transition pathways are distinguished
by the solid and dashed arrows. There is one saddle point for
connecting any two stable node points. \label{fig3}}
\end{center}
\end{figure*}

In order to clearly exhibit the transition process at different
regions, in Fig.\ref{fig3} we give the contour plots of the
effective potential $V$ for four representative points $(\mu,
\sigma)$ marked by stars in Fig.\ref{fig2}. In the region $n_s=9$,
there are two possible transition pathways, each of which contains a
two-step transition process via an intermediate metastable state.
Let $\Delta V_1^{(\alpha)}$ and $\Delta V_2^{(\alpha)}$ denote the
energy barrier of the first step and the second step for the
$\alpha$-th transition pathway ($\alpha=1,2$), respectively. The
corresponding transition rates are
\begin{eqnarray}
{R^{(\alpha )}} = \frac{1}{{\frac{1}{{R_1^{(\alpha )}}} +
\frac{1}{{R_2^{(\alpha )}}}}}, \label{eq4}
\end{eqnarray}
with
\begin{eqnarray}
R_{1,2}^{(\alpha )} = Z\exp ( - {{\Delta V_{1,2}^{(\alpha )}}
\mathord{\left/
 {\vphantom {{\Delta V_{1,2}^{(\alpha )}} D}} \right.
 \kern-\nulldelimiterspace} D}). \label{eq5}
\end{eqnarray}
Here the prefactor $Z$ is given by \cite{PhysRev.121.1668}
\begin{eqnarray}
Z = {{\omega _n^{(1)}\omega _n^{(2)} {\omega _s^{(1)}} } \over {2\pi
\omega _s^{(2)}}}. \label{eq6}
\end{eqnarray}
Here $\omega _s^{(1,2)} = \sqrt {\left| {e_s^{(1,2)}} \right|}$ and
$\omega _n^{(1,2)} = \sqrt {\left| {e_n^{(1,2)}} \right|}$ are the
vibrational frequencies at the saddle point and at the node point
from which the system escapes, where ${e_s^{(1,2)}}$ ($e_s^{(1)}<0$)
and ${e_n^{(1,2)}}$ are the eigenvalues of Hessian matrix at the
saddle point and at the node point, respectively,

The total transition rate is
\begin{eqnarray}
R=p^{(1)} R^{(1)}+p^{(2)} R^{(2)}, \label{eq7}
\end{eqnarray}
where
\begin{eqnarray}
{p^{(1,2)}} = \frac{{R_1^{^{(1,2)}}}}{{R_1^{^{(1)}} + R_1^{^{(2)}}}}
\label{eq8}
\end{eqnarray}
are the probabilities of the two transition pathways happening.

In the region $n_s=7$, there are also two possible transition
pathways. However, one pathway also contains a two-step process, but
the other one becomes a one-step process. Let $\Delta V_1^{(1)}$ and
$\Delta V_2^{(1)}$ denote the energy barrier of the first step and
the second step for the two-step transition pathway, respectively,
and $\Delta V^{(2)}$ the energy barrier for the one-step transition
pathway. The total transition rate is also expressed by
Eq.\ref{eq7}, but we have
\begin{eqnarray}
\left\{ \begin{array}{l} {R^{(2)}} = Z\exp ( - {{\Delta {V^{(2)}}}
\mathord{\left/
 {\vphantom {{\Delta {V^{(2)}}} D}} \right.
 \kern-\nulldelimiterspace} D})\\
{p^{(1)}} = \frac{{R_1^{^{(1)}}}}{{R_1^{^{(1)}} + {R^{(2)}}}}\\
{p^{(2)}} = \frac{{{R^{(2)}}}}{{R_1^{^{(1)}} + {R^{(2)}}}}
\end{array} \right. \label{eq9}
\end{eqnarray}

In the region $n_s=5$, there are also two possible transition
pathways, but both of them contain a one-step process. Let $\Delta
V^{(1)}$ and $\Delta V^{(2)}$ denote the energy barrier of the two
transition pathways. The terms in the total transition rate by
Eq.\ref{eq7} become
\begin{eqnarray}
\left\{ \begin{array}{l} {R^{(1,2)}} = Z\exp ( - {{\Delta
{V^{(1,2)}}} \mathord{\left/
 {\vphantom {{\Delta {V^{(1,2)}}} D}} \right.
 \kern-\nulldelimiterspace} D})\\
{p^{(1,2)}} = \frac{{{R^{(1,2)}}}}{{{R^{(1)}} + {R^{(2)}}}}
\end{array} \right. \label{eq10}
\end{eqnarray}

Note that for the case $\sigma=0$, the two oscillators are identical
and the two possible transition pathways are equivalent. However, as
$\sigma$ is increased, the probability of the nucleation pathway
marked by the solid line in Fig.\ref{fig3} quickly approaches one
due to a larger force acted on the second oscillator, such that the
total transition rate is almost determined by this dominant
transition pathway.

Lastly, in the region $n_s=3$ there is a single one-step transition
pathway. The nucleation rate is $R = Z\exp ( - {{\Delta V}
\mathord{\left/  {\vphantom {{\Delta V} D}} \right.
 \kern-\nulldelimiterspace} D})$, where $\Delta V$ is the energy barrier
of the transition process.

\begin{figure*}
\begin{center}
\includegraphics [width=12cm]{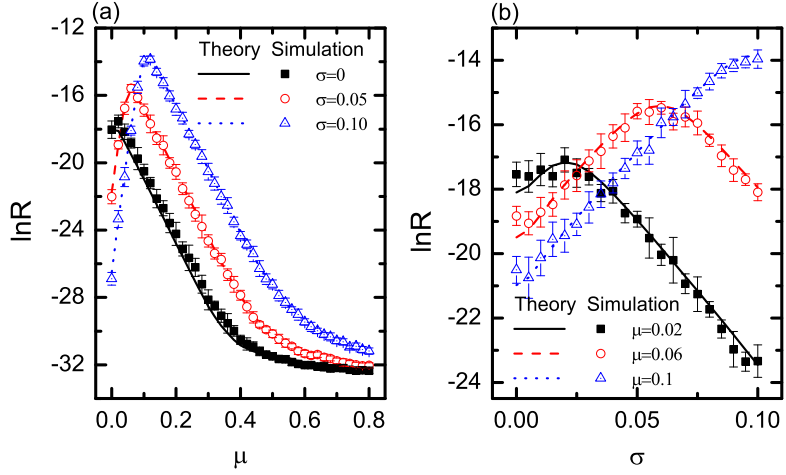}
\caption{(color online) The logarithm of the total transition rate
$\ln R$ as a function of $\mu$ for different $\sigma$ (a), and as a
function of $\sigma$ for different $\mu$ (b). The theoretical
results and FFS simulation ones are indicated by lines and symbols,
respectively. The noise intensity is fixed at $D=0.01$.
\label{fig4}}
\end{center}
\end{figure*}

So far, we have theoretically obtained the pathway and the rate of
the transition in various parameters regions. For identical
oscillators ($\sigma=0$), the system undergoes a transition from a
two-step transition process with two possible pathways (transition
in a serial way) to a one-step process with also two possible
pathways at $\mu=\mu_1=\mu_2$ (transition in a synchronized way) to
a one-step process with a single pathway at $\mu=\mu_3$. For
nonidentical oscillators ($\sigma\neq0$), a new transition way
emerges at $\mu_1<\mu<\mu_2$ that is a candidate of a two-step
pathway and a one-step pathway. In a word, in both cases there
exists a critical coupling in which the transition changes from a
two-step process to a one-step one. Interestingly, similar
phenomenon was also reported in disturbed coupled nonlinear
oscillators with purely deterministic dynamics
\cite{PNAS2006,EPL2007,APPB2008,PhysRevE.81.026603}.

To validate the theory, we have performed extensive numerical
simulations for Langevin equation \ref{eq1}. However, the transition
is an activated process that occurs extremely slow, and brute-force
simulation is thus prohibitively expensive. To overcome this
difficulty, we will use a recently developed simulation method,
forward flux sampling (FFS) \cite{PRL05018104, JPH09463102}. All the
simulation results are obtained via averaging over 20 independent
FFS simulations.

In Fig.\ref{fig4}(a), we show that the logarithm of the total
transition rate $\ln R$ as a function of $\mu$ for different
$\sigma$: $0$, $0.05$, and $0.1$. The theoretical results and FFS
simulation ones are indicated by lines and symbols, respectively.
There are an excellent agreement between them. For $\sigma=0$, $\ln
R$ slightly increases and then decreases monotonically as $\mu$ is
increased. Interestingly, if $\sigma$ becomes larger, for example
$\sigma=0.5$, $\ln R$ clearly exhibits a nonmonotonic dependence on
$\sigma$: $\ln R$ peaks at $\mu\simeq0.06$. With further increasing
$\sigma$, such a peak becomes more clear and the location of the
peak shifts to a larger $\mu$. In Fig.\ref{fig4}(b), we show that
the dependence of $\ln R$ on $\sigma$ at different $\mu$. Clearly,
$\ln R$ behaves significantly different dependence trend with
$\sigma$ for different $\mu$. For $\mu<0.1$, $\ln R$ depends
nonmonotonically on $\sigma$. There exists a maximal rate at a
moderate magnitude of force disorder $\sigma$. On the other hand,
for $\mu\geq0.1$, $\ln R$ increases monotonically with $\sigma$.
This implies that in the presence of a weak coupling a proper level
of force mismatch can enhance the occurrence of the transition
process. While for a relatively strong coupling, a larger magnitude
of force mismatch is favorable to advance the process.

\begin{figure*}
\begin{center}
\includegraphics [width=15cm]{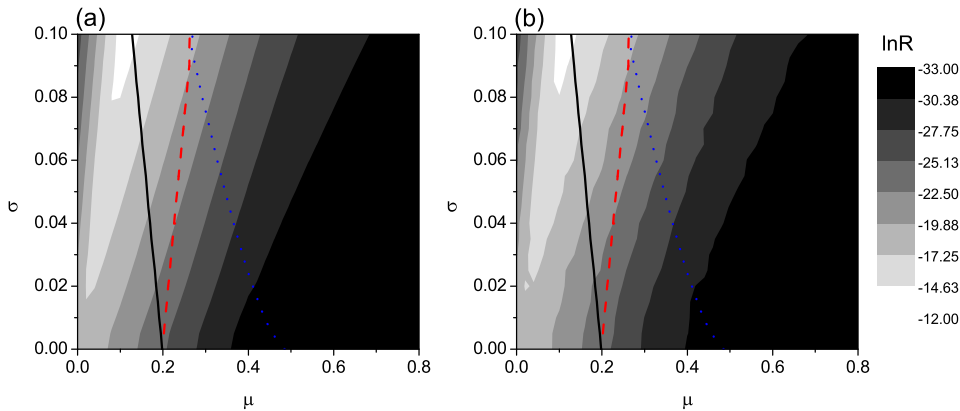}
\caption{(color online) The contour plot of $\ln R$ in the
$\mu-\sigma$ plane obtained by both theoretical calculation (a) and
FFS simulations (b). The lines mark the separatrix as shown in
Fig.2. The other parameter is $D=0.01$. \label{fig5}}
\end{center}
\end{figure*}

Furthermore, we summarize the results of $\ln R$ in the plane of
parametric $\mu-\sigma$, as shown in Fig.\ref{fig5}. The results are
obtained both by theoretical calculation and by FFS simulations that
are in good agreements. The region in which $\ln R$ is globally
maximal appears near $(\mu, \sigma) \simeq(0.1, 0.1)$, and the
region $\ln R$ is globally minimal locates at the right bottom of
the $\mu-\sigma$ plane. That is to say, if one wants to globally
accelerate transition process, the force mismatch should be maximal
while the coupling needs to be delicately chosen. If one intends to
suppress the occurrence of transition, the force mismatch should be
minimal and the coupling should be maximal.

\section{Conclusions}
To conclude, we have studied the transition process in a system of
two mutually coupling forced bistable oscillators. By constructing
the effective potential, we have shown that how the pathway and the
rate of the transition change with the coupling strength and the
force mismatch. We have identified four types of transition process
at different parametric regions: (i) two possible transition
pathways each containing a two-step transition process; (ii) two
possible transition pathways one containing a two-step transition
process and the other containing a one-step transition process;
(iii) two possible transition pathways each containing a one-step
transition process; (iv) a single one-step transition pathway.
Furthermore, the total transition rate shows a rich dependence on
coupling and force mismatch. On the one hand, as the coupling is
strengthened the rate increases and then decreases, i.e., a maximal
rate exists for an intermediate magnitude of coupling. On the other
hand, the rate also shows a nonmonotonic dependence on the force
mismatch for a weak coupling. While for a relatively strong
coupling, the rate increases monotonically with the force mismatch.
All the results have been validated to be in an excellent agreement
with the extensive FFS simulations.

Obviously, partial results of the present work can give some
valuable understanding for some existing phenomena reported
previously
\cite{PhysRevLett.109.138104,PhysRevLett.112.148101,PhysRevE.83.046124,arXiv2014,PNAS2006}.
Meanwhile, our findings may provide potential implications for
controlling transition events in coupled systems by a delicately
chosen coupling or parametric mismatch. In the further work, the
investigation of the more complex activated transition process in
more than two coupled oscillators is desirable.

\begin{acknowledgments}
We acknowledge supports from the National Science Foundation of
China (11205002, 11475003, 21125313), ¡°211 project¡± of Anhui
University (02303319-33190133), and Anhui Provincial Natural Science
Foundation (1408085MA09).
\end{acknowledgments}


\begin{thebibliography}{34}
\expandafter\ifx\csname
natexlab\endcsname\relax\def\natexlab#1{#1}\fi
\expandafter\ifx\csname bibnamefont\endcsname\relax
  \def\bibnamefont#1{#1}\fi
\expandafter\ifx\csname bibfnamefont\endcsname\relax
  \def\bibfnamefont#1{#1}\fi
\expandafter\ifx\csname citenamefont\endcsname\relax
  \def\citenamefont#1{#1}\fi
\expandafter\ifx\csname url\endcsname\relax
  \def\url#1{\texttt{#1}}\fi
\expandafter\ifx\csname urlprefix\endcsname\relax\def\urlprefix{URL
}\fi \providecommand{\bibinfo}[2]{#2}
\providecommand{\eprint}[2][]{\url{#2}}

\bibitem[{\citenamefont{Wernsdorfer et~al.}(1997)\citenamefont{Wernsdorfer,
  Hasselbach, Benoit, Barbara, Doudin, Meier, Ansermet, and
  Mailly}}]{PhysRevB.55.11552}
\bibinfo{author}{\bibfnamefont{W.}~\bibnamefont{Wernsdorfer}},
  \bibinfo{author}{\bibfnamefont{K.}~\bibnamefont{Hasselbach}},
  \bibinfo{author}{\bibfnamefont{A.}~\bibnamefont{Benoit}},
  \bibinfo{author}{\bibfnamefont{B.}~\bibnamefont{Barbara}},
  \bibinfo{author}{\bibfnamefont{B.}~\bibnamefont{Doudin}},
  \bibinfo{author}{\bibfnamefont{J.}~\bibnamefont{Meier}},
  \bibinfo{author}{\bibfnamefont{J.-P.} \bibnamefont{Ansermet}},
  \bibnamefont{and} \bibinfo{author}{\bibfnamefont{D.}~\bibnamefont{Mailly}},
  \bibinfo{journal}{Phys. Rev. B} \textbf{\bibinfo{volume}{55}},
  \bibinfo{pages}{11552} (\bibinfo{year}{1997}).

\bibitem[{\citenamefont{Fulton and Dunkleberger}(1974)}]{PhysRevB.9.4760}
\bibinfo{author}{\bibfnamefont{T.~A.} \bibnamefont{Fulton}} \bibnamefont{and}
  \bibinfo{author}{\bibfnamefont{L.~N.} \bibnamefont{Dunkleberger}},
  \bibinfo{journal}{Phys. Rev. B} \textbf{\bibinfo{volume}{9}},
  \bibinfo{pages}{4760} (\bibinfo{year}{1974}).

\bibitem[{\citenamefont{Kashchiev}(2000)}]{Kashchiev2000}
\bibinfo{author}{\bibfnamefont{D.}~\bibnamefont{Kashchiev}},
  {\bibinfo{title}{Nucleation: basic theory with applications}}
  (\bibinfo{publisher}{Butterworths-Heinemann}, \bibinfo{address}{Oxford},
  \bibinfo{year}{2000}).

\bibitem[{\citenamefont{Auer and Frenkel}(2004)}]{Frenkel2004}
\bibinfo{author}{\bibfnamefont{S.}~\bibnamefont{Auer}} \bibnamefont{and}
  \bibinfo{author}{\bibfnamefont{D.}~\bibnamefont{Frenkel}},
  \bibinfo{journal}{Ann. Rev. Phys. Chem.} \textbf{\bibinfo{volume}{55}},
  \bibinfo{pages}{333} (\bibinfo{year}{2004}).

\bibitem[{\citenamefont{Gillespie}(1977)}]{Gillespie1977}
\bibinfo{author}{\bibfnamefont{D.~T.} \bibnamefont{Gillespie}},
  \bibinfo{journal}{J. Chem. Phys} \textbf{\bibinfo{volume}{81}},
  \bibinfo{pages}{2340} (\bibinfo{year}{1977}).

\bibitem[{\citenamefont{van Kampen}(1992)}]{Kampen1992}
\bibinfo{author}{\bibfnamefont{N.~G.} \bibnamefont{van Kampen}},
  {\bibinfo{title}{Stochastic Processes in Physics and Chemistry}}
  (\bibinfo{publisher}{Elsevier}, \bibinfo{address}{Amsterdam},
  \bibinfo{year}{1992}).

\bibitem[{\citenamefont{A.~Sali and Karplus}(1994)}]{Karplus1994}
\bibinfo{author}{\bibfnamefont{E.~S.} \bibnamefont{A.~Sali}} \bibnamefont{and}
  \bibinfo{author}{\bibfnamefont{M.}~\bibnamefont{Karplus}},
  \bibinfo{journal}{Nature} \textbf{\bibinfo{volume}{369}},
  \bibinfo{pages}{248} (\bibinfo{year}{1994}).

\bibitem[{\citenamefont{White and Wimley}(1999)}]{White1999}
\bibinfo{author}{\bibfnamefont{S.~H.} \bibnamefont{White}} \bibnamefont{and}
  \bibinfo{author}{\bibfnamefont{W.~C.} \bibnamefont{Wimley}},
  \bibinfo{journal}{Annu. Rev. Biophys. Biomol. Struct.}
  \textbf{\bibinfo{volume}{28}}, \bibinfo{pages}{319} (\bibinfo{year}{1999}).

\bibitem[{\citenamefont{Wales}(2003)}]{Wales2003}
\bibinfo{author}{\bibfnamefont{D.}~\bibnamefont{Wales}},
  {\bibinfo{title}{Energy Landscapes: Applications to Clusters,
  Biomolecules and Glasses}} (\bibinfo{publisher}{Cambridge University Press},
  \bibinfo{address}{Cambridge, England}, \bibinfo{year}{2003}).

\bibitem[{\citenamefont{Assaf and Meerson}(2006)}]{PhysRevLett.97.200602}
\bibinfo{author}{\bibfnamefont{M.}~\bibnamefont{Assaf}} \bibnamefont{and}
  \bibinfo{author}{\bibfnamefont{B.}~\bibnamefont{Meerson}},
  \bibinfo{journal}{Phys. Rev. Lett.} \textbf{\bibinfo{volume}{97}},
  \bibinfo{pages}{200602} (\bibinfo{year}{2006}).

\bibitem[{\citenamefont{Dykman et~al.}(2008)\citenamefont{Dykman, Schwartz, and
  Landsman}}]{PhysRevLett.101.078101}
\bibinfo{author}{\bibfnamefont{M.~I.} \bibnamefont{Dykman}},
  \bibinfo{author}{\bibfnamefont{I.~B.} \bibnamefont{Schwartz}},
  \bibnamefont{and} \bibinfo{author}{\bibfnamefont{A.~S.}
  \bibnamefont{Landsman}}, \bibinfo{journal}{Phys. Rev. Lett.}
  \textbf{\bibinfo{volume}{101}}, \bibinfo{pages}{078101}
  (\bibinfo{year}{2008}).

\bibitem[{\citenamefont{Kramers}(1940)}]{Kramers1940}
\bibinfo{author}{\bibfnamefont{H.}~\bibnamefont{Kramers}},
  \bibinfo{journal}{Physica (Utrecht)} \textbf{\bibinfo{volume}{7}},
  \bibinfo{pages}{284} (\bibinfo{year}{1940}).

\bibitem[{\citenamefont{H\"anggi et~al.}(1990)\citenamefont{H\"anggi, Talkner,
  and Borkovec}}]{RevModPhys.62.251}
\bibinfo{author}{\bibfnamefont{P.}~\bibnamefont{H\"anggi}},
  \bibinfo{author}{\bibfnamefont{P.}~\bibnamefont{Talkner}}, \bibnamefont{and}
  \bibinfo{author}{\bibfnamefont{M.}~\bibnamefont{Borkovec}},
  \bibinfo{journal}{Rev. Mod. Phys.} \textbf{\bibinfo{volume}{62}},
  \bibinfo{pages}{251} (\bibinfo{year}{1990}).

\bibitem[{\citenamefont{Sung and Park}(1996)}]{PhysRevLett.77.783}
\bibinfo{author}{\bibfnamefont{W.}~\bibnamefont{Sung}} \bibnamefont{and}
  \bibinfo{author}{\bibfnamefont{P.~J.} \bibnamefont{Park}},
  \bibinfo{journal}{Phys. Rev. Lett.} \textbf{\bibinfo{volume}{77}},
  \bibinfo{pages}{783} (\bibinfo{year}{1996}).

\bibitem[{\citenamefont{Sebastian and Paul}(2000)}]{PhysRevE.62.927}
\bibinfo{author}{\bibfnamefont{K.~L.} \bibnamefont{Sebastian}}
  \bibnamefont{and} \bibinfo{author}{\bibfnamefont{A.~K.~R.}
  \bibnamefont{Paul}}, \bibinfo{journal}{Phys. Rev. E}
  \textbf{\bibinfo{volume}{62}}, \bibinfo{pages}{927} (\bibinfo{year}{2000}).

\bibitem[{\citenamefont{Kraikivski et~al.}(2004)\citenamefont{Kraikivski,
  Lipowsky, and Kierfeld}}]{EPL2004}
\bibinfo{author}{\bibfnamefont{P.}~\bibnamefont{Kraikivski}},
  \bibinfo{author}{\bibfnamefont{R.}~\bibnamefont{Lipowsky}}, \bibnamefont{and}
  \bibinfo{author}{\bibfnamefont{J.}~\bibnamefont{Kierfeld}},
  \bibinfo{journal}{Europhys. Lett.} \textbf{\bibinfo{volume}{66}},
  \bibinfo{pages}{763} (\bibinfo{year}{2004}).

\bibitem[{\citenamefont{L.~I.~McCann and Golding}(1999)}]{NatureDykman}
\bibinfo{author}{\bibfnamefont{M.~D.} \bibnamefont{L.~I.~McCann}}
  \bibnamefont{and} \bibinfo{author}{\bibfnamefont{B.}~\bibnamefont{Golding}},
  \bibinfo{journal}{Nature} \textbf{\bibinfo{volume}{402}},
  \bibinfo{pages}{785} (\bibinfo{year}{1999}).

\bibitem[{\citenamefont{Wiesenfeld et~al.}(1996)\citenamefont{Wiesenfeld,
  Colet, and Strogatz}}]{PhysRevLett.76.404}
\bibinfo{author}{\bibfnamefont{K.}~\bibnamefont{Wiesenfeld}},
  \bibinfo{author}{\bibfnamefont{P.}~\bibnamefont{Colet}}, \bibnamefont{and}
  \bibinfo{author}{\bibfnamefont{S.~H.} \bibnamefont{Strogatz}},
  \bibinfo{journal}{Phys. Rev. Lett.} \textbf{\bibinfo{volume}{76}},
  \bibinfo{pages}{404} (\bibinfo{year}{1996}).

\bibitem[{\citenamefont{A.E.~Motter and Nishikawa}(2013)}]{Motter2013}
\bibinfo{author}{\bibfnamefont{M.~A.} \bibnamefont{A.E.~Motter},
  \bibfnamefont{S.A.~Myers}} \bibnamefont{and}
  \bibinfo{author}{\bibfnamefont{T.}~\bibnamefont{Nishikawa}},
  \bibinfo{journal}{Nat. Phys.} \textbf{\bibinfo{volume}{9}},
  \bibinfo{pages}{191} (\bibinfo{year}{2013}).

\bibitem[{\citenamefont{Bar-Yam and Epstein}(2004)}]{PNAS04004341}
\bibinfo{author}{\bibfnamefont{Y.}~\bibnamefont{Bar-Yam}} \bibnamefont{and}
  \bibinfo{author}{\bibfnamefont{I.~R.} \bibnamefont{Epstein}},
  \bibinfo{journal}{Proc. Natl. Acad. Sci. USA} \textbf{\bibinfo{volume}{101}},
  \bibinfo{pages}{4341} (\bibinfo{year}{2004}).

\bibitem[{\citenamefont{Tian and Burrage}(2006)}]{PNAS06008372}
\bibinfo{author}{\bibfnamefont{T.}~\bibnamefont{Tian}} \bibnamefont{and}
  \bibinfo{author}{\bibfnamefont{K.}~\bibnamefont{Burrage}},
  \bibinfo{journal}{Proc. Natl. Acad. Sci. USA} \textbf{\bibinfo{volume}{103}},
  \bibinfo{pages}{8372} (\bibinfo{year}{2006}).

\bibitem[{\citenamefont{Koseska et~al.}(2009)\citenamefont{Koseska, Zaikin,
  Kurths, and Garc\'ia-Ojalvo}}]{Plos09004872}
\bibinfo{author}{\bibfnamefont{A.}~\bibnamefont{Koseska}},
  \bibinfo{author}{\bibfnamefont{A.}~\bibnamefont{Zaikin}},
  \bibinfo{author}{\bibfnamefont{J.}~\bibnamefont{Kurths}}, \bibnamefont{and}
  \bibinfo{author}{\bibfnamefont{J.}~\bibnamefont{Garc\'ia-Ojalvo}},
  \bibinfo{journal}{PLoS ONE} \textbf{\bibinfo{volume}{4}},
  \bibinfo{pages}{e4872} (\bibinfo{year}{2009}).

\bibitem[{\citenamefont{Khasin et~al.}(2012{\natexlab{a}})\citenamefont{Khasin,
  Meerson, Khain, and Sander}}]{PhysRevLett.109.138104}
\bibinfo{author}{\bibfnamefont{M.}~\bibnamefont{Khasin}},
  \bibinfo{author}{\bibfnamefont{B.}~\bibnamefont{Meerson}},
  \bibinfo{author}{\bibfnamefont{E.}~\bibnamefont{Khain}}, \bibnamefont{and}
  \bibinfo{author}{\bibfnamefont{L.~M.} \bibnamefont{Sander}},
  \bibinfo{journal}{Phys. Rev. Lett.} \textbf{\bibinfo{volume}{109}},
  \bibinfo{pages}{138104} (\bibinfo{year}{2012}{\natexlab{a}}).

\bibitem[{\citenamefont{Khasin et~al.}(2012{\natexlab{b}})\citenamefont{Khasin,
  Khain, and Sander}}]{PhysRevLett.109.248102}
\bibinfo{author}{\bibfnamefont{M.}~\bibnamefont{Khasin}},
  \bibinfo{author}{\bibfnamefont{E.}~\bibnamefont{Khain}}, \bibnamefont{and}
  \bibinfo{author}{\bibfnamefont{L.~M.} \bibnamefont{Sander}},
  \bibinfo{journal}{Phys. Rev. Lett.} \textbf{\bibinfo{volume}{109}},
  \bibinfo{pages}{248102} (\bibinfo{year}{2012}{\natexlab{b}}).

\bibitem[{\citenamefont{Lombardo et~al.}(2014)\citenamefont{Lombardo, Gambassi,
  and Dall'Asta}}]{PhysRevLett.112.148101}
\bibinfo{author}{\bibfnamefont{P.}~\bibnamefont{Lombardo}},
  \bibinfo{author}{\bibfnamefont{A.}~\bibnamefont{Gambassi}}, \bibnamefont{and}
  \bibinfo{author}{\bibfnamefont{L.}~\bibnamefont{Dall'Asta}},
  \bibinfo{journal}{Phys. Rev. Lett.} \textbf{\bibinfo{volume}{112}},
  \bibinfo{pages}{148101} (\bibinfo{year}{2014}).

\bibitem[{\citenamefont{Chen and Hou}(2011)}]{PhysRevE.83.046124}
\bibinfo{author}{\bibfnamefont{H.}~\bibnamefont{Chen}} \bibnamefont{and}
  \bibinfo{author}{\bibfnamefont{Z.}~\bibnamefont{Hou}},
  \bibinfo{journal}{Phys. Rev. E} \textbf{\bibinfo{volume}{83}},
  \bibinfo{pages}{046124} (\bibinfo{year}{2011}).

\bibitem[{\citenamefont{A.~Nematzadeh and Ahn}(2014)}]{arXiv2014}
\bibinfo{author}{\bibfnamefont{A.~F.} \bibnamefont{A.~Nematzadeh},
  \bibfnamefont{E.~Ferrara}} \bibnamefont{and}
  \bibinfo{author}{\bibfnamefont{Y.-Y.} \bibnamefont{Ahn}},
  \bibinfo{journal}{arXiv:1401.1257}  (\bibinfo{year}{2014}).

\bibitem[{\citenamefont{Landauer and Swanson}(1961)}]{PhysRev.121.1668}
\bibinfo{author}{\bibfnamefont{R.}~\bibnamefont{Landauer}} \bibnamefont{and}
  \bibinfo{author}{\bibfnamefont{J.~A.} \bibnamefont{Swanson}},
  \bibinfo{journal}{Phys. Rev.} \textbf{\bibinfo{volume}{121}},
  \bibinfo{pages}{1668} (\bibinfo{year}{1961}).

\bibitem[{\citenamefont{Mezi\ifmmode~\acute{c}\else
  \'{c}\fi{}}(2006)}]{PNAS2006}
\bibinfo{author}{\bibfnamefont{I.}~\bibnamefont{Mezi\ifmmode~\acute{c}\else
  \'{c}\fi{}}}, \bibinfo{journal}{Proc. Natl. Acad. Sci. U.S.A.}
  \textbf{\bibinfo{volume}{103}}, \bibinfo{pages}{7542} (\bibinfo{year}{2006}).

\bibitem[{\citenamefont{Hennig et~al.}(2007)\citenamefont{Hennig,
  Schimansky-Geier, and H\"anggi}}]{EPL2007}
\bibinfo{author}{\bibfnamefont{D.}~\bibnamefont{Hennig}},
  \bibinfo{author}{\bibfnamefont{L.}~\bibnamefont{Schimansky-Geier}},
  \bibnamefont{and} \bibinfo{author}{\bibfnamefont{P.}~\bibnamefont{H\"anggi}},
  \bibinfo{journal}{Europhys. Lett.} \textbf{\bibinfo{volume}{78}},
  \bibinfo{pages}{20002} (\bibinfo{year}{2007}).

\bibitem[{\citenamefont{H\"anggi et~al.}(2008)\citenamefont{H\"anggi, Fugmann,
  and Schimansky-Geier}}]{APPB2008}
\bibinfo{author}{\bibfnamefont{P.}~\bibnamefont{H\"anggi}},
  \bibinfo{author}{\bibfnamefont{S.}~\bibnamefont{Fugmann}}, \bibnamefont{and}
  \bibinfo{author}{\bibfnamefont{L.}~\bibnamefont{Schimansky-Geier}},
  \bibinfo{journal}{Acta Phys. Pol. B} \textbf{\bibinfo{volume}{39}},
  \bibinfo{pages}{1125} (\bibinfo{year}{2008}).

\bibitem[{\citenamefont{Eisenhower and Mezi\ifmmode~\acute{c}\else
  \'{c}\fi{}}(2010)}]{PhysRevE.81.026603}
\bibinfo{author}{\bibfnamefont{B.}~\bibnamefont{Eisenhower}} \bibnamefont{and}
  \bibinfo{author}{\bibfnamefont{I.}~\bibnamefont{Mezi\ifmmode~\acute{c}\else
  \'{c}\fi{}}}, \bibinfo{journal}{Phys. Rev. E} \textbf{\bibinfo{volume}{81}},
  \bibinfo{pages}{026603} (\bibinfo{year}{2010}).

\bibitem[{\citenamefont{Allen et~al.}(2005)\citenamefont{Allen, Warren, and ten
  Wolde}}]{PRL05018104}
\bibinfo{author}{\bibfnamefont{R.~J.} \bibnamefont{Allen}},
  \bibinfo{author}{\bibfnamefont{P.~B.} \bibnamefont{Warren}},
  \bibnamefont{and} \bibinfo{author}{\bibfnamefont{P.~R.} \bibnamefont{ten
  Wolde}}, \bibinfo{journal}{Phy. Rev. Lett.} \textbf{\bibinfo{volume}{94}},
  \bibinfo{pages}{018104} (\bibinfo{year}{2005}).

\bibitem[{\citenamefont{Allen et~al.}(2009)\citenamefont{Allen, Valeriani, and
  ten Wolde}}]{JPH09463102}
\bibinfo{author}{\bibfnamefont{R.~J.} \bibnamefont{Allen}},
  \bibinfo{author}{\bibfnamefont{C.}~\bibnamefont{Valeriani}},
  \bibnamefont{and} \bibinfo{author}{\bibfnamefont{P.~R.} \bibnamefont{ten
  Wolde}}, \bibinfo{journal}{J. Phys.: Condens. Matter}
  \textbf{\bibinfo{volume}{21}}, \bibinfo{pages}{463102}
  (\bibinfo{year}{2009}).

\end{thebibliography}

\end{document}